\documentclass[aps,prl,twocolumn,preprintnumbers,superscriptaddress,10pt,longbibliography]{revtex4-2}

\usepackage[T1]{fontenc}
\usepackage[utf8]{inputenc}

\usepackage{amssymb}
\usepackage{amsmath}
\usepackage{amsfonts}
\usepackage{graphicx}
\usepackage{subfigure}
\usepackage{color}
\usepackage{dcolumn}
\usepackage{hyperref}

\usepackage{tikz}
\usetikzlibrary{arrows.meta,decorations.markings}

\newcommand{\eq}{\begin{equation}}
\newcommand{\eqx}{\end{equation}}
\newcommand{\eqn}{\begin{eqnarray}}
\newcommand{\eqnx}{\end{eqnarray}}
\newcommand{\be}{\begin{equation}}
\newcommand{\ee}{\end{equation}}
\newcommand{\bea}{\begin{eqnarray}}
\newcommand{\eea}{\end{eqnarray}}

\newcommand{\f}[2]{\frac{#1}{#2}}

\newcommand{\nn}{{\mathcal N}}

\newcommand{\eps}{\varepsilon}

\newcommand{\sg}{\sigma}

\newcommand{\dl}{\delta}
\newcommand{\Dl}{\Delta}
\renewcommand{\th}{\theta}
\newcommand{\om}{\omega}
\newcommand{\Om}{\Omega}

\newcommand{\Gm}{\Gamma}

\newcommand{\cor}[1]{\left\langle{#1}\right\rangle}
\newcommand{\OO}[1]{{\mathcal O}\left(#1\right)}

\newcommand{\Tr}{\operatorname{Tr}}
\newcommand{\sgn}{\operatorname{sign}}
\newcommand{\Reo}{\Re e}
\newcommand{\Imo}{\Im m}

\DeclareMathOperator{\tr}{tr}

\newcommand{\QNM}{\mathrm{Q N M}}

\begin{document}

\title{Quasi-Normal Modes and thermalization in integrable Quantum Field Theories\\ 
from the Thermodynamic Bethe Ansatz}

\author{Zoltan Bajnok}
\email{bajnok.zoltan@wigner.hu}
\affiliation{HUN-REN Wigner Research Centre for Physics,
1121 Budapest, Konkoly-Thege Miklós út 29-33, Hungary}
\affiliation{Institute of Theoretical Physics and Mark Kac Center for Complex Systems Research, Jagiellonian University, Łojasiewicza 11, 30-348  Kraków, Poland}
\author{Romuald A. Janik}
\email{romuald.janik@gmail.com}
\affiliation{Institute of Theoretical Physics and Mark Kac Center for Complex Systems Research, Jagiellonian University, Łojasiewicza 11, 30-348  Kraków, Poland}

\begin{abstract}
We propose a Thermodynamic Bethe Ansatz (TBA) approach for describing thermalization in integrable quantum field theories. The exponential damping and oscillations when approaching thermal equilibrium can be understood as the analogs of quasi-normal mode (QNM) frequencies of black hole perturbations within the framework of holography.
Holographic intuitions and some general properties of thermal Green's functions motivate the existence of such a TBA formulation, as well as
lead to the imposition of a nonstandard analytical structure on the QNM TBA, while comparison with integrable form-factor expansions fixes the remaining ingredients. We conjecture the QNM TBA to be equivalent to performing an all-order resummation of secular terms in a thermal form-factor expansion of retarded 2-point correlation functions. 
We compare the predictions with tensor network simulations of the E8 deformation of Ising CFT at nonzero temperature.
We conjecture that the QNM TBA framework should also be applicable for describing out-of-equilibrium approach to Generalized Hydrodynamics (GHD).
\end{abstract}

\maketitle


\noindent {\bf Introduction.} The process of thermalization, describing the approach to thermal equilibrium, is one of the most basic and fundamental properties of complex physical systems. It is physically interesting and nontrivial even in its linearized form, when one considers perturbations away from thermal equilibrium and the nature of their relaxation to the thermal state. 

Here, one has to distinguish perturbations associated with (slowly varying) conserved charges, which are associated with hydrodynamics (or Generalized Hydrodynamics (GHD) for integrable QFTs) from genuinely non-equilibrium thermalization processes. The former are driven by spacetime gradients of conserved currents and are quite well understood~\cite{Castro_Alvaredo_2016,Bertini2016,Doyon_2020,Delacr_taz_2022,Davison_2025}, while the latter remain elusive and will be the focus of the present paper.

We will mainly concentrate on the large real-time exponential behaviour of the \emph{integrated} retarded correlator
\eq
\f{i}{2\pi} \th(t) \cor{\int dx\, [O(t,x), O(0,0)]}_T \!\!\!\sim A(t) \cdot e^{-\Gm t -i \Omega t} +const
\label{e.retcorr}
\eqx
where $A(t)$ is sub-exponential.
This zero momentum ($k=0$) case is of particular interest, as it 
does not excite any time-dependent GHD modes due to its spatial uniformity and thus involves genuinely non-equilibrium effects.

Despite the apparently simple formulation, the problem of determining the exponential thermal damping describing the approach to thermal equilibrium remains as an outstanding challenge despite many years of investigation~\cite{DamleSachdev1998,LeClairMussardo1999,Saleur2000,EsslerKonik2008,EsslerKonik2009,PozsgayTakacs2010,Doyon2005,CuberoPanfil2019,KormosVorosZarand2022}.
Indeed, even for very well understood integrable QFTs, the problem of computing real-time thermal damping becomes extremely difficult. 

When extracting the damping coefficient of a thermal 2-point function, one would have to use infinite form-factor expansions at finite temperature in the timelike regime, isolate secular terms arising from kinematical singularities and resum them into an exponential. 
Basically the only known exact result was obtained by such a \textit{tour-de-force} computation in ~\cite{Granet_2020}, where the damping coefficient of the $\sg$ operator in a thermally deformed Ising CFT was obtained \footnote{Below we quote the variant for the ray along $x=0$, which will be relevant for the $k=0$ case considered in the present paper.} to be
\eq
\label{e.essler1}
\Gm_{\mathrm{GFE}} = \f{1}{2\pi} \int_{-\infty}^\infty dp\, \f{|p|}{E_p} \log \coth \f{E_p}{2T}
\eqx

\begin{figure}[t]
    \centering
    \resizebox{\linewidth}{!}{
\begin{tikzpicture}[
  x=1cm,y=1cm,
  font=\normalsize,
  line cap=round,line join=round,
  cylinder/.style={draw=black,line width=0.85pt},
  sheet/.style={draw=black,line width=0.75pt},
  sk hidden/.style={draw=black,line width=0.55pt,
    dash pattern=on 2.1pt off 2.1pt},
  sk direction/.style={draw=black,line width=0.75pt,
    -{Stealth[length=1.7mm,width=1.1mm]}},
  flow/.style={draw=black,line width=0.95pt,
    -{Stealth[length=3.0mm,width=2.0mm]}},
  continuation/.style={draw=black,line width=0.95pt,
    <->,>={Stealth[length=2.8mm,width=1.9mm]}},
  equation/.style={inner sep=0pt,outer sep=0pt},
  thermal/.style={inner sep=0pt,outer sep=0pt}
]

\def\EuclideanLabel{%
  G_E^{+}(n,k)=\frac{B_E^{+}(n,k)}{A_E^{+}(n,k)},\quad n>0}
\def\TransferLabel{%
  A_E^{+}\!\left(n,i\Delta E_{\mathrm{cyl}}\right)=0,\quad n>0}
\def\QNMLabel{%
  A_E^{+}\!\left(-\frac{i\omega_{\mathrm{QNM}}}{2\pi T},k\right)=0}
\def\HamiltonianLabel{H_{\mathrm{cyl}}}
\def\ThermalLabel{\dfrac{1}{T}}
\def\ContinuationLabel{Analytical continuation}

\def\CylinderLength{3.10}
\def\CylinderRadius{0.61}
\def\HorizontalRimDepth{0.22}
\def\VerticalRadius{0.59}
\def\VerticalRimDepth{0.19}
\def\VerticalHalfLength{1.30}
\def\SheetLength{\CylinderLength}
\def\SheetHeight{1.69}
\def\SKBackAngle{80} 
\def\SKFrontAngle{45} 
\def\SKFoldRise{0.24}

\coordinate (EuclideanOrigin) at (0.35,6.45);
\coordinate (TransferOrigin)  at (10.20,6.45);
\coordinate (LorentzianOrigin) at (0.35,0.76);
\coordinate (EuclideanEquation) at (2.50,4.83);
\coordinate (TransferEquation) at (10.25,4.45);
\coordinate (QNMEquation) at (5.13,1.2); 

\begin{scope}[shift={(EuclideanOrigin)}]
  \draw[cylinder]
    (0,\CylinderRadius) -- (\CylinderLength,\CylinderRadius);
  \draw[cylinder]
    (0,-\CylinderRadius) -- (\CylinderLength,-\CylinderRadius);
  \draw[cylinder]
    (0,0) ellipse[x radius=\HorizontalRimDepth,
                  y radius=\CylinderRadius];
  \draw[cylinder]
    (\CylinderLength,\CylinderRadius)
    arc[start angle=90,end angle=-90,
        x radius=\HorizontalRimDepth,y radius=\CylinderRadius];
  \node[thermal,anchor=west] at ({\CylinderLength+0.52},0)
    {$\ThermalLabel$};
\end{scope}
\node[equation] at (EuclideanEquation)
  {$\displaystyle\EuclideanLabel$};

\draw[flow] (5.37,6.45) -- (8.15,6.45);

\begin{scope}[shift={(TransferOrigin)}]
  \draw[cylinder]
    (-\VerticalRadius,\VerticalHalfLength)
    -- (-\VerticalRadius,-\VerticalHalfLength);
  \draw[cylinder]
    (\VerticalRadius,\VerticalHalfLength)
    -- (\VerticalRadius,-\VerticalHalfLength);
  \draw[cylinder]
    (0,\VerticalHalfLength)
    ellipse[x radius=\VerticalRadius,y radius=\VerticalRimDepth];
  \draw[cylinder]
    (-\VerticalRadius,-\VerticalHalfLength)
    arc[start angle=180,end angle=360,
        x radius=\VerticalRadius,y radius=\VerticalRimDepth];
  \draw[flow] (-0.35,-0.46) -- (-0.35,0.36);
  \node[inner sep=0pt,anchor=west] at (-0.19,-0.09)
    {$\HamiltonianLabel$};
  \node[thermal,anchor=west]
    at ({\VerticalRadius+0.24},{-\VerticalHalfLength-0.0}) 
    {$\ThermalLabel$};
\end{scope}
\node[equation] at (TransferEquation)
  {$\displaystyle\TransferLabel$};

\draw[flow] (2, 4.1 ) -- (2.,3.3);

\begin{scope}[shift={(LorentzianOrigin)}]
  \coordinate (SKNearBackBank) at
    ({\HorizontalRimDepth*cos(\SKBackAngle)},
     {\CylinderRadius*sin(\SKBackAngle)});
  \coordinate (SKNearFrontBank) at
    ({\HorizontalRimDepth*cos(\SKFrontAngle)},
     {\CylinderRadius*sin(\SKFrontAngle)});
  \path (SKNearBackBank) ++(\SheetLength,0)
    coordinate (SKFarBackBank);
  \path (SKNearFrontBank) ++(\SheetLength,0)
    coordinate (SKFarFrontBank);
  \path (SKNearBackBank) ++(0,\SheetHeight)
    coordinate (SKNearBackTop);
  \path (SKNearFrontBank) ++(0,\SheetHeight)
    coordinate (SKNearFrontTop);
  \path (SKFarBackBank) ++(0,\SheetHeight)
    coordinate (SKFarBackTop);
  \path (SKFarFrontBank) ++(0,\SheetHeight)
    coordinate (SKFarFrontTop);

  \draw[sk hidden]
    (SKFarBackBank)
    arc[start angle=\SKBackAngle,end angle=270,
        x radius=\HorizontalRimDepth,y radius=\CylinderRadius];
  \draw[cylinder]
    (\SheetLength,-\CylinderRadius)
    arc[start angle=270,end angle={360+\SKFrontAngle},
        x radius=\HorizontalRimDepth,y radius=\CylinderRadius];
  \draw[cylinder]
    (0,-\CylinderRadius) -- (\SheetLength,-\CylinderRadius);
  \draw[cylinder]
    (SKNearBackBank)
    arc[start angle=\SKBackAngle,end angle={360+\SKFrontAngle},
        x radius=\HorizontalRimDepth,y radius=\CylinderRadius];

  \draw[cylinder] (SKNearFrontBank) -- (SKFarFrontBank);

  \draw[sheet]
    (SKNearBackBank) -- (SKNearBackTop)
    .. controls +(0,\SKFoldRise) and +(0,\SKFoldRise)
    .. (SKNearFrontTop) -- (SKNearFrontBank);

\draw[sk hidden]
  (SKFarBackBank) -- (SKFarBackTop);

\draw[sheet]
  (SKFarBackTop)
  .. controls +(0,\SKFoldRise) and +(0,\SKFoldRise)
  .. (SKFarFrontTop) -- (SKFarFrontBank);  
  
\draw[sk hidden]
  (SKFarBackBank) -- (SKFarBackTop);

\draw[sheet]
  (SKFarBackTop)
  .. controls +(0,\SKFoldRise) and +(0,\SKFoldRise)
  .. (SKFarFrontTop) -- (SKFarFrontBank);
  

  

\pgfmathsetmacro{\SKFoldDeltaY}{
  \CylinderRadius*(sin(\SKFrontAngle)-sin(\SKBackAngle))
}
\pgfmathsetmacro{\SKFoldTopT}{
  \SKFoldRise / (
    \SKFoldRise-\SKFoldDeltaY
    +sqrt((\SKFoldRise)^2+(\SKFoldDeltaY)^2)
  )
}

\path
  (SKNearBackTop)
  .. controls +(0,\SKFoldRise) and +(0,\SKFoldRise)
  .. (SKNearFrontTop)
  coordinate[pos=\SKFoldTopT] (SKNearFoldTop);

\path
  (SKFarBackTop)
  .. controls +(0,\SKFoldRise) and +(0,\SKFoldRise)
  .. (SKFarFrontTop)
  coordinate[pos=\SKFoldTopT] (SKFarFoldTop);

\draw[sheet] (SKNearFoldTop) -- (SKFarFoldTop);


  \draw[sk direction] (SKNearBackBank) ++(0,0.34) -- ++(0,0.62);
  \draw[sk direction] (SKNearFrontBank) ++(0,1.35) -- ++(0,-0.62);

\draw[sk hidden]
  (SKNearBackBank) -- (SKFarBackBank);

  \coordinate (SheetPoint) at (0.62+1,{\CylinderRadius+0.8}); 
    \node[equation] at (SheetPoint) {$\displaystyle G_R(\om, k)$};  
  \node[thermal,anchor=west] at ({\SheetLength+0.52},0)
    {$\ThermalLabel$};
\end{scope}
\node[equation,anchor=west] at (QNMEquation)
  {$\displaystyle\QNMLabel$};

\coordinate (ContinuationLower) at (7.30,1.96);
\coordinate (ContinuationControlLower) at (8.21,2.12);
\coordinate (ContinuationControlUpper) at (9.15,3.14);
\coordinate (ContinuationUpper) at (9.48,3.91);
\draw[continuation]
  (ContinuationLower)
  .. controls (ContinuationControlLower) and (ContinuationControlUpper)
  .. (ContinuationUpper);
\node[anchor=west,inner sep=0pt,font=\normalsize]
  at (8.60,2.38) {\ContinuationLabel}; 

\end{tikzpicture}}
    \caption{Relations between Euclidean Green's function, Lorentzian retarded thermal correlator and the spectral problem for the theory on a Lorentzian cylinder of size $1/T$. 
    The energies $\Dl E_{cyl}$ are measured w.r.t. cylinder ground state. 
    This indicates an analytically continued spectral sheet from positive Matsubara modes on a Lorentzian cylinder, leading to the postulated TBA description of QNMs and thus thermal damping.}
    \label{fig.overview}
\end{figure}
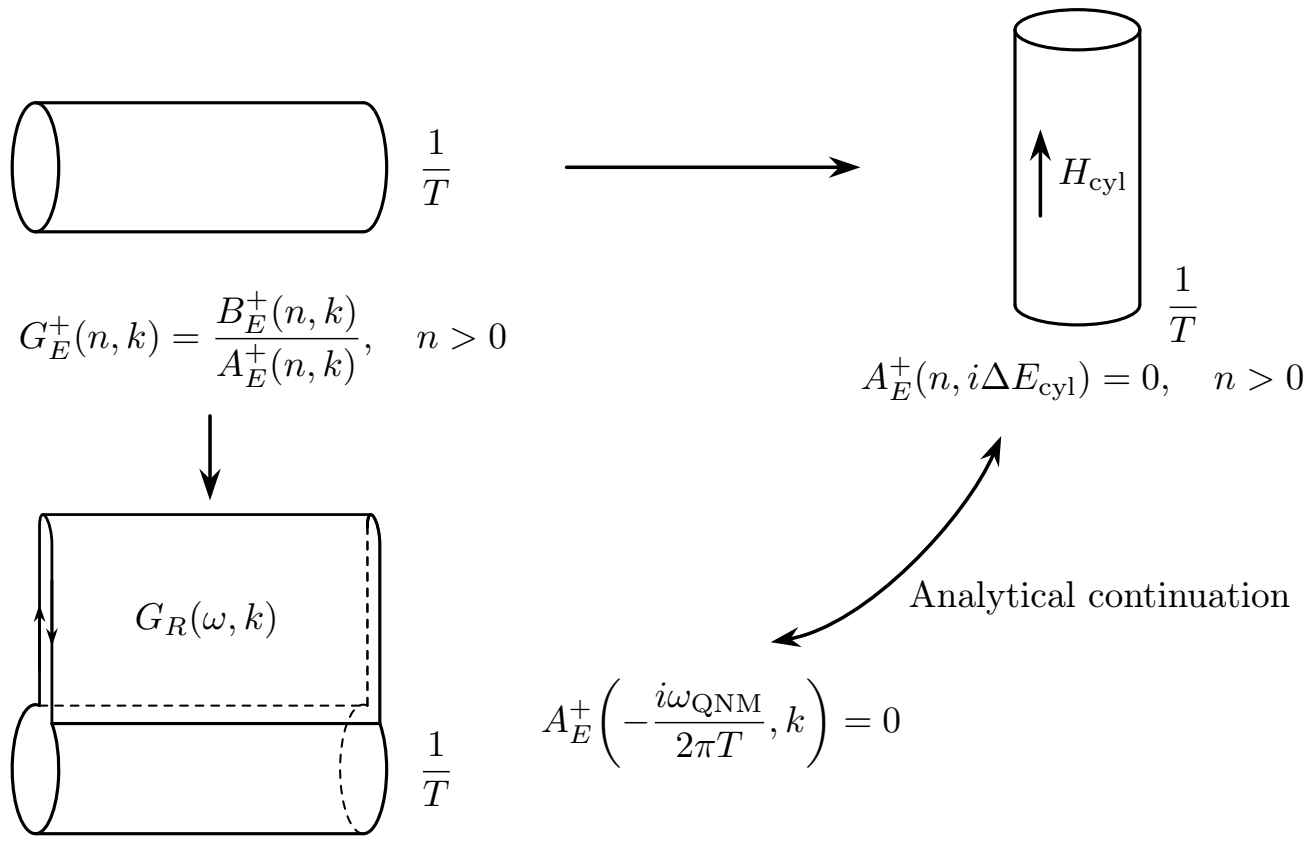

The key claim of the present paper is that the thermal damping coefficient and the oscillation frequency 
\eq
\om_{\QNM} = - i \Gm + \Omega ,
\eqx
which are combined into a complex Quasi-Normal-Mode (QNM) frequency,
can be obtained from a (modified) Thermodynamic Bethe Ansatz~\cite{Zamolodchikov1990TBA} formulation, which does not involve \emph{a-posteriori} any considerations of form-factors but just the infinite volume scattering data of the integrable QFT (see Fig.~\ref{fig.overview} for an overview). 
This QNM TBA thus has to automatically resum all secular terms in the thermal form factor expansion and is conjectured to provide a universal framework for addressing thermal damping in integrable QFTs.

Apart from direct physical applications, the process of thermalization is also very interesting from a quite different perspective. Indeed, for theories which posses a holographic dual description through the AdS/CFT correspondence~\cite{Maldacena_1999,Witten:1998qj}, the linearized thermalization process gets mapped into perturbations of  dual black holes (representing thermal equilibrium) --- the so-called quasi-normal modes (QNM)~\cite{Berti_2009}, which exhibit exponential decay due to ingoing boundary conditions at the black hole horizon. The QNMs correspond to poles of the finite temperature retarded 2-point functions in the boundary theory in the energy-momentum $\omega/k$ representation (see e.g.~\cite{Kovtun:2005ev}). This relation can be made completely explicit for 2D conformal field theories, where the exponential decays of thermal perturbations are in 1:1 correspondence with QNMs of a three-dimensional BTZ black hole~\cite{Banados:1992wn}
\eq
\label{e.cftqnm}
\omega^{{\mathrm{CFT/BTZ}}}_{\QNM} = -i\, 2 \pi T \left(\Dl +2p \right) +|k| \quad \text{for}\quad p=0, 1,2,\ldots
\eqx
where $\Dl$ is the scaling dimension of the corresponding primary operator~\cite{Birmingham:2001pj,Son:2002sd}.
This holographic gravitational point of view is also another source of motivation for the investigation of how these QNMs behave when we move away from the conformal point by turning on an integrable massive deformation. 
Moreover, the holographic picture and the relation between Matsubara modes in the Euclidean black hole and Lorentzian QNMs were also a key inspiration for formulating our main hypothesis.

We argue that the existence and general form of the QNM TBA can be motivated on physical grounds (see Fig.~\ref{fig.overview} for an overview), and constrained by the known exact data on CFTs and the thermally deformed Ising CFT. These arguments, however, do not completely determine additional potential driving terms which could appear in a \emph{massive interacting} integrable TBA.

To fix these terms, we analyze a sector of secular terms in the thermal form-factor expansion of~\cite{Pozsgay:2018tns} and show that they are consistent with a QNM TBA of the above form with specific additional driving terms. 
We cross-check the numerical predictions with a tensor network simulation of the E8 theory.
We conclude the paper with some additional conjectures and a discussion.


\medskip

\noindent {\bf The spectral problem for QNMs.}  
When analyzing form-factor expansions for thermal 2-point functions, their long time behaviour may seem \emph{a-priori} inextricably intertwined with the details of the form-factors of the relevant operators. The holographic picture provides, however, a quite different perspective. There, the operators in the QFT correspond to bulk fields, whose propagation in a Lorentzian black hole background leads to a set of complex QNM frequencies. Various retarded 2-point functions of boundary operators then essentially couple to this set of modes (in an operator dependent way selected by non-vanishing form factors). This suggests that one should be able to obtain a universal spectrum of QNMs irrespective of the details of the form-factors. Indeed, we will observe below exactly this phenomenon, when considering secular terms in the integrable form-factor expansion.

It is instructive to remain still within the holographic picture and compare the near-boundary expansion of QNMs in a Lorentzian black hole
\eq
\label{e.ingoing}
\phi_{\mathrm{ingoing}}(\om, k, z) \sim A(\om, k) z^{\f{2 -\Dl}{2}} + B(\om, k) z^{\f{\Dl}{2}} + \ldots
\eqx
with Matsubara modes in an Euclidean black hole
\eq
\phi_{\mathrm{Matsubara}}(n, k, z) \sim A_E(n, k) z^{\f{2 -\Dl}{2}} + B_E(n, k) z^{\f{\Dl}{2}} + \ldots
\eqx
The condition for QNM frequencies is $A(\om_{\QNM}, k)=0$, which coincides with a pole of the retarded Green's function $G_R(\om,k) \propto B(\om, k)/A(\om, k)$. The vanishing of the leading coefficient $A(\om_{\QNM}, k)=0$ in~\eqref{e.ingoing}, also has the interpretation of normalizability condition of the fluctuation in the bulk spacetime, which identifies physical modes.
The pole condition $A(\om_{\QNM}, k)=0$ can be obtained by analytically continuing from \emph{positive} Matsubara modes $A_E^+(n, k)$, $n>0$ as $n \to -i\om_{\QNM}/(2\pi T)$~\cite{Denef:2009kn}.

Clearly, all this can be also directly stated in thermal QFT as the retarded Green's function can be obtained from the Euclidean one by analytical continuation from positive Matsubara modes.
On the other hand, the poles of the Euclidean Green's function for integer $n$ and purely imaginary $k=\pm i \Dl E_{\mathrm{\mathrm{cyl}}}$, can be identified with physical states for a theory on a Lorentzian cylinder of radius $r=1/T$, where $\Dl E_{cyl}$ is the energy over the cylinder ground state (see Fig.~\ref{fig.overview}).

The above discussion suggests that QNM frequencies and physical states on a cylinder coexist on some complex spectral surface analytically continued from \emph{positive} integer momenta on a cylinder, as illustrated in Fig.~\ref{fig.overview}:
\eq
A_E^+ \left( n \to -\f{i\om_{\QNM}}{2\pi T}, k\right) = 0 \ 
\longleftrightarrow
\ 
A_E^+ \left(n, i \Dl E_{cyl}\right) = 0
\eqx
The right hand side condition (i.e. the spectrum on the cylinder) is clearly given by standard TBA equations for the given 2D integrable QFT. It is thus very natural to expect that there should be some form of analytically continued TBA formulation (more generally, specific analytic structure imposed on the Y-system) which would determine the complex QNM frequencies 
for an integrable QFT. 
Explicitly, $k$~and $\om_{\QNM}= -i \Gm + \Omega$ are related to the energy and momentum of the analytically continued TBA by
\eq
\label{e.gapqnm}
\widetilde{\Dl E_{\mathrm{cyl}}} \equiv \widetilde{E} - \widetilde{E_{\mathrm{gs}}} = \pm i\, k 
\quad 
\widetilde{\Dl P_{\mathrm{cyl}}} \equiv \widetilde{P} - \widetilde{P_{\mathrm{gs}}} = -i\, \om_{\QNM}
\eqx
so we are always dealing with a gap, hence a pair of TBA solutions.

The discussion in terms of physicality conditions $A(\om_{\QNM}, k)=0$ instead of the Green's functions is suggestive here, as this would be a direct counterpart of TBA equations which \emph{do not} compute a Green's function, and hence miss the numerator $B(\om, k)$ which is, however, irrelevant for our purposes.

Let us note, that there exists already a nontrivial analytically continued spectral problem in the context of $\nn=4$ Super-Yang-Mills, namely the relation between the spectrum (operator dimension in terms of spin) and the intercept of the BFKL pomeron (spin in terms of conformal weight). There, the change required nontrivial modifications in the analytical conditions for the Y-system/Quantum Spectral Curve~\cite{Janik:2013nqa,Alfimov:2014bwa}. 
We expect that the Y-system/TBA for QNMs will also be characterized by some nonstandard analytical structure.


\medskip

\noindent {\bf The general form of QNM TBA.} As it seems to be very difficult to directly perform the analytical continuation from the spectral problem at \emph{positive} momenta, we will adapt a different strategy and try to guess the analytical requirements based on the general properties identified above and the available known exact results. We focus on the zero momentum $k=0$ case, which describes the behaviour of the integrated correlator~\eqref{e.retcorr}.
As $k=0$ in the spectral problem represents vanishing energy, we are fixing the TBA energy to zero and the QNM frequency will be given by the momentum (analytically continued $2\pi T\, n$)
through~\eqref{e.gapqnm}.

Consider first the case of a 2D CFT and a scalar primary operator. A descendant with positive momentum on the cylinder will have conformal weights $(h+p+n, h+p)$ with $n>0$. Since for a 2D CFT, the left moving and right moving chiral TBAs are completely separate (with driving terms $e^{\pm \theta}$), this suggests that analytically continuing from positive $n$ should only move the right moving wing, leaving the left moving solution unchanged.

Since we are interested in a zero energy state, the right moving ``QNM state'' should be a ``hole'' state with opposite charges to the left-moving one i.e. $(-h-p, h+p)$. Then the computed momentum indeed reproduces the correct CFT damping
\eq
\label{e.Gmcft}
\Gm_{\mathrm{CFT}} = 2\pi T \left(\Dl + 2p\right)
\eqx
Note that in the above there is no Casimir term $(-c/24,-c/24)$, so we really compute the difference of the momentum between the ``QNM state'' and the continued vacuum (see~\eqref{e.gapqnm}).
The state to ``hole'' transformation can be realized on the level of the (chiral) TBA by an overall shift $\th \to \th+i \pi$, which exactly reverses all charges including both the contributions of the active singularities as well as the integrated part.

When going away from the conformal point, the left and right chiral parts are no longer decoupled and hence have to be joined together into a single solution of the Y-system \footnote{For concreteness, we sometimes refer to the SLYM one, $Y(\th+ia)Y(\th-ia)=1+Y(\th)$ with $a=\f{\pi}{3}$, and denote the corresponding shift operator by $D_af(\th)=f(\th+ia)+f(\th-i a)$.}.

This can be done in two ways. The most obvious one is to choose a \textit{tanh}-like contour interpolating between $Im\, \th=0$ and $Im\, \th=\pi$. This requires, however, a $m/T \sinh\th$ driving term (in order for $e^{-\eps(\th)}$ to vanish at infinity at both ends of the contour), which turns out to be incompatible with the result~\eqref{e.essler1}, which requires the free solution to be $\eps(\th)=\pm m/T\cosh\th$, as we will see below. 

\begin{figure}[t]
    \centering
    \resizebox{\linewidth}{!}{
\begin{tikzpicture}[
  x=1cm,y=1cm,
  font=\normalsize,
  line cap=round,line join=round,
  f2 axis/.style={line width=0.45pt,-{Latex[length=2.2mm,width=1.6mm]}},
  f2 imaginary axis/.style={
    f2 axis,dash pattern=on 3.6pt off 4pt
  },
  f2 contour/.style={
    line width=1.15pt,
    postaction={decorate},
    decoration={markings,
      mark=at position 0.62 with {
        \arrow{Latex[length=3.1mm,width=2.2mm]}
      }
    }
  }
]
  \def\PanelOffset{4.30}      
  \def\RealHalfWidth{2.60}    
  \def\AxisBottom{-2.00}
  \def\AxisTop{2.45}
  \def\ShiftHeight{1.20}     
  \def\PositiveContourEnd{2.38}

  \begin{scope}[shift={(-\PanelOffset,0)}]
    \coordinate (theta-origin) at (0,0);
    \coordinate (theta-left) at (-\RealHalfWidth,0);
    \coordinate (theta-right) at (\RealHalfWidth,0);
    \coordinate (theta-shift-start) at (0,\ShiftHeight);
    \coordinate (theta-shift-end) at (\PositiveContourEnd,\ShiftHeight);

    \draw[f2 axis] (theta-left) -- (theta-right);
    \draw[f2 imaginary axis] (0,\AxisBottom) -- (0,\AxisTop);
    \node[below right,inner sep=2pt] at (theta-right) {$\theta$};
    \node[left,inner sep=3pt] at (theta-shift-start) {$i\pi$};

    \draw[f2 contour] (theta-left) -- (theta-origin);
    \draw[f2 contour] (theta-shift-start) -- (theta-shift-end);
  \end{scope}

  \node[font=\Huge,scale=1.15,inner sep=0pt]
    at (0,0) {$\Longrightarrow$};

  \begin{scope}[shift={(\PanelOffset,0)}]
    \coordinate (x-origin) at (0,0);
    \coordinate (x-left) at (-\RealHalfWidth,0);
    \coordinate (x-right) at (\RealHalfWidth,0);
    \coordinate (x-contour-end) at (\PositiveContourEnd,0);

    \draw[f2 axis] (x-left) -- (x-right);
    \draw[f2 imaginary axis] (0,\AxisBottom) -- (0,\AxisTop);
    \node[below right,inner sep=2pt] at (x-right) {$y$};

    \draw[f2 contour] (x-left) -- (x-origin);
    \draw[f2 contour] (x-origin) -- (x-contour-end);
  \end{scope}
\end{tikzpicture}}
    \caption{QNM TBA contours in the original rapidity variable $\th$ (left) and in the $y$ variable equal to $\th$ for $\th<0$ and equal to $\th-i\pi$ for $\th>0$ (right). The dashed imaginary axis is a cut discontinuity of the Y-system/TBA.}
    \label{fig.contours}
\end{figure}
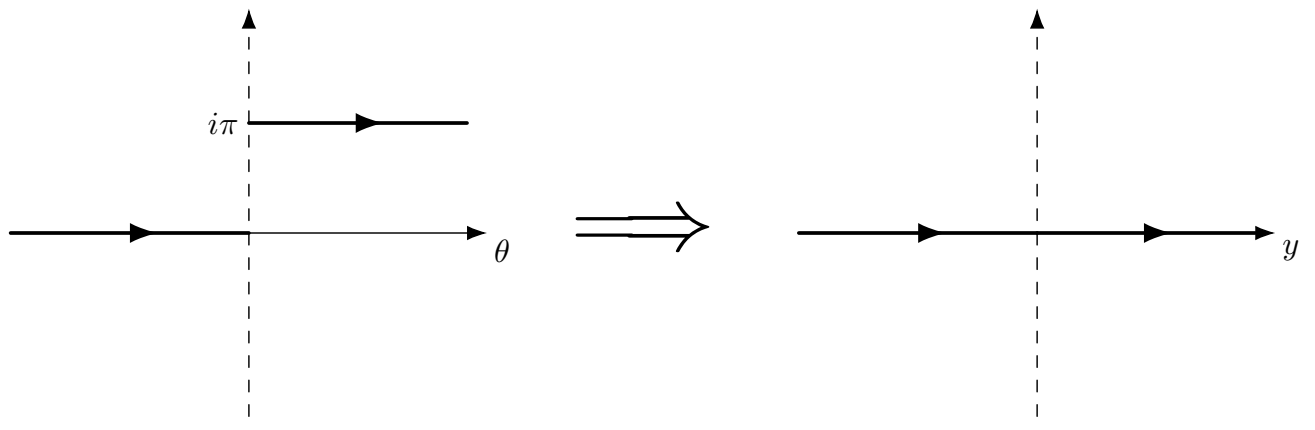

In order to be able to accommodate a \textit{cosh}-type small~$T$ solution preserving the correct behaviour at infinity, the driving term should have a discontinuity like
\eq
\label{e.drivingtheta}
-\f{m}{T} \cdot \mathrm{sign}(\Re e\, \th) \cdot \cosh \th
\eqx
This leads to 
the other option, namely introducing a cut along the imaginary axis and gluing together a TBA with $\Im m\, \th=0$ on the left and $\Im m\, \th=\pi$ on the right (see Fig.~\ref{fig.contours}) with the above discontinuous driving term~\eqref{e.drivingtheta}. Remarkably enough, this is exactly the structure indicated by the form factor expansion discussed below.

Introducing the $y$ variable as in Fig.~\ref{fig.contours}(right), this yields the following form of the TBA equations for the continued ground state and the ``QNM state'':
\eqn
\label{e.tba1gen}
\eps_0 &=& \f{m}{T}\cosh y - J\star L_0+ d_0(y) \\
\eps_1 &=& \f{m}{T}\cosh y - J\star L_1 + d_1(y)
\label{e.tba2gen}
\eqnx
where $L_i=\log\left(1+e^{-\eps_i}\right)$, 
and we added some unspecified, for the moment, additional driving terms satisfying appropriate conditions like $(D_a-1) d_i=0$ in the SLYM case. The $J$ kernel is related to the standard TBA $K$ kernel by
\eq
\label{e.J}
J(y_1, y_2) = \mathrm{sign}(y_1) K(y_1-y_2) \mathrm{sign}(y_2)
\eqx
and satisfies the same functional equation as $K$. 
It is important to note that due to the shift by $i\pi$ on the right, the integrated part of the expression for momentum $P_1 - P_0$
involves the \emph{absolute value} of $\sinh y$:
\eq
\label{e.tbamomentum1}
\int_{-\infty}^\infty \f{dy}{2\pi} |m\sinh y| (L_1(y)-L_0(y))
\eqx
In the case of the thermal deformation of the Ising model, the kernels vanish, and the TBA solutions are simply $\eps_0(y)=m/T\cosh y$ and $\eps_1(y)=i\pi + m/T\cosh y$, representing the NS and R vacua. The above formula~\eqref{e.tbamomentum1} then reproduces exactly the result~\eqref{e.essler1} of~\cite{Granet_2020}. 
In the following section, we will analyze a form-factor expansion of a thermal 2-point correlation function and deduce the driving terms $d_i(y)$, as well as cross-checking part of the structure of~\eqref{e.tba1gen}-\eqref{e.tba2gen}. 


\medskip

\noindent {\bf Form factor expansion and the  QNM TBA.} 
The best way to give a meaning to the thermal two point function is the finite volume regularization~\cite{PozsgayTakacs2008a,PozsgayTakacs2008b} 
\begin{equation}
G(x,t)=\lim_{R\to\infty}\frac{\mathrm{Tr}_{R}
 [e^{-{H_R}/{T}}O(0,0)O(x,t)]}{\mathrm{Tr}_{R}[e^{-{H_R}/{T}}]}
 \label{eq:defGR}
\end{equation}
By inserting a complete set of finite $R$-volume states and form factors and carefully cancelling $R$-extensive terms the limit is well-defined for any $x$ and $t$  \cite{Szecsenyi:2012jq}. For space-like separations, $|x|>t$, a partial resummation leads to the Pozsgay-Szécsényi (PSz) formula~\cite{Pozsgay:2018tns} 
\begin{equation}
 G(x,t)
 =\sum_{n=0}^{\infty}\frac1{n!}
 \int\prod_{j=1}^{n}\frac{d\theta_j}{2\pi}\,f(\theta_j)
 \mathcal F_{n,c}(\theta_1,\ldots,\theta_n;x,t)
 \label{PSz}
\end{equation}
where $f$ is the thermal filling,  
which comes from the TBA 
\eq
\varepsilon_0
=
\frac{m}{T} \cosh\th -
K \star L_0,
\qquad
f=\frac{1}{1+e^{\varepsilon_0}}.
\label{eps0}
\eqx 
and ${\mathcal F}_{n,c}$ is the connected form factor of the bilocal operator. 

The large $x>0$ behaviour (for $t=0$) can be easily understood from the mirror formulation where $r=1/T$ serves as the volume of the infinite cylinder and the two-point function can be expanded in terms of finite volume form factors 
\eq
G(x,0)=\sum_{\vert n\rangle_r} {}_r\langle{0}\vert O\vert n\rangle_r\, {}_r \langle{n}\vert O\vert 0\rangle_r \,e^{-x\Delta E_n } 
\eqx
For large (Euclidean) times, $x$, it is dominated by a standing one particle state whose TBA is  
\eq
\varepsilon_1
=
\frac{m}{T} \cosh\th -\log s - 
K \star L_1,
\label{eps1}
\eqx
where the source term $s$ depends on the model. For the sinh-Gordon theory it is $s(\theta)= S(\theta + i \pi/2)$ and the gap takes the form
\eq
\Delta E_1=m-m\int \frac{d\theta}{2\pi } \cosh \theta \ (L_1-L_0)
\label{E1}
\eqx
The asymptotic behaviour 
\eq
G(x,0)= {}_r\langle{0}\vert O\vert 1\rangle_r\, {}_r \langle{1}\vert O\vert 0\rangle_r e^{-x\Delta E_1 }+\dots  
\eqx
can also be extracted in principle from the PSz formula. One needs to resum the secular terms dominating the large $x$ behaviour. They come from the definition of the connected form factors as well as from integration domains with form factor singularities. Due to the recursive structure of the form factor singularities the contributions have to be resummed by the non-linear integral equation
\begin{equation}
L_1-L_0=\log( 1-f+f s e^{K\star(L_1-L_0)})
\label{gap}
\end{equation}
This  is the gap equation, which follows from taking the difference of the TBA equations~\eqref{eps1} and~\eqref{eps0}. 

Let us now switch on $t>0$ and introduce $\xi=x/t$. In the construction of the connected form factors  intermediate states are inserted and their rapidity is integrated over. The phase corresponding to such a particle is 
\(
 e^{i t(E(u)-\xi p(u))}
\)
and the contour is shifted downwards the real line $u\to u-i\eta$. As far as $x>t$ the real part of the exponent is negative, the integrals are convergent and are evaluated typically at $\eta=\pi/2$. In particular $s$ appears from this shift as $s(\theta)=S(\theta+i \pi/2)$.  If, however, we continue $t$ larger than $|x|$, then the exponent changes sign at
$ u_\xi =\operatorname{arctanh}\xi $ so we have to deform the contour partially upwards  for $u>u_\xi$. In doing so we cross the kinematical singularities of the form factors, whose residue terms we have to pick up. These lower order form factor terms can be regrouped, which leads to the time-like generalization of PSz
(see \textit{Supplemental Material}). 
For the integrated correlator the relevant limit is when $\xi=0$ i.e. $u_\xi =0$.  One can then show that the effect of these terms is changing the kernel and the source terms as 
\eq
K\to J=\Sigma K \Sigma \ ; \quad s \to c \ ,\ c(\theta)=S(-\vert \theta \vert)   
\label{KtoJ}
\eqx
where $\Sigma g(\theta)=\mathrm{sign}(\theta) g(\theta)$, so we recover~\eqref{e.J} and obtain the modified source.  
This changes the gap equation \eqref{gap} for $\chi=L_1-L_0$ to 
\begin{equation}
\chi
=
\log\!\left[
1-f+
f c e^{J\star \chi}
\right]
\label{main}
\end{equation} This continued integral equation can be rewritten into a difference form~\eqref{e.tba1gen}-\eqref{e.tba2gen} with individual sources 
\eq
\label{e.sources}
d_0=(J-K)\star L_0\quad  d_1= - \log c + (J-K)\star L_0
\eqx
The driving term $d_0$ ensures that the thermal groundstate is not changed, while $d_1$ has additionally the continuation of the source.  The effect of the contour shift and saddle point integration leads to 
the damping rate reproducing~\eqref{e.tbamomentum1}
\eq
-\Gamma+i\delta m=
m\int \frac{dy}{2\pi}\,
\vert \sinh(y)\vert \chi(y)
\label{Gammam}
\eqx
For the case of the Ising model, $J$ vanishes and \(\chi=\log(1-2f)=\log\tanh( E/2T)\). Thus  \eqref{main} and~\eqref{Gammam} directly reproduce the Ising damping exponent~\eqref{e.essler1}.

\begin{figure}[t]
\includegraphics[width=2.5cm]{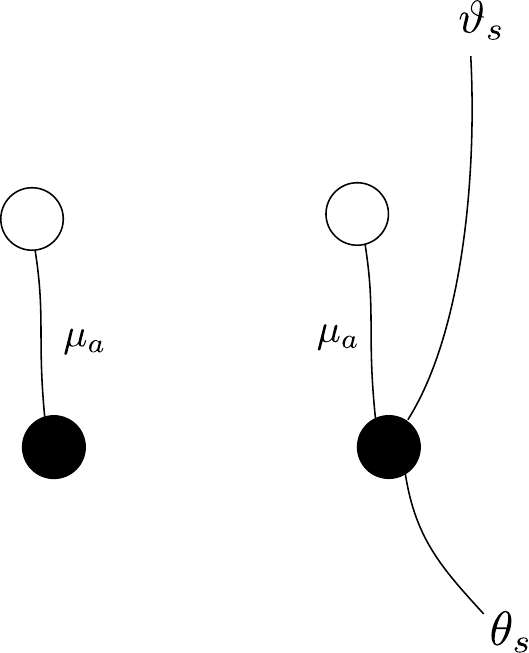}\hspace{1.0cm}\includegraphics[width=4cm]{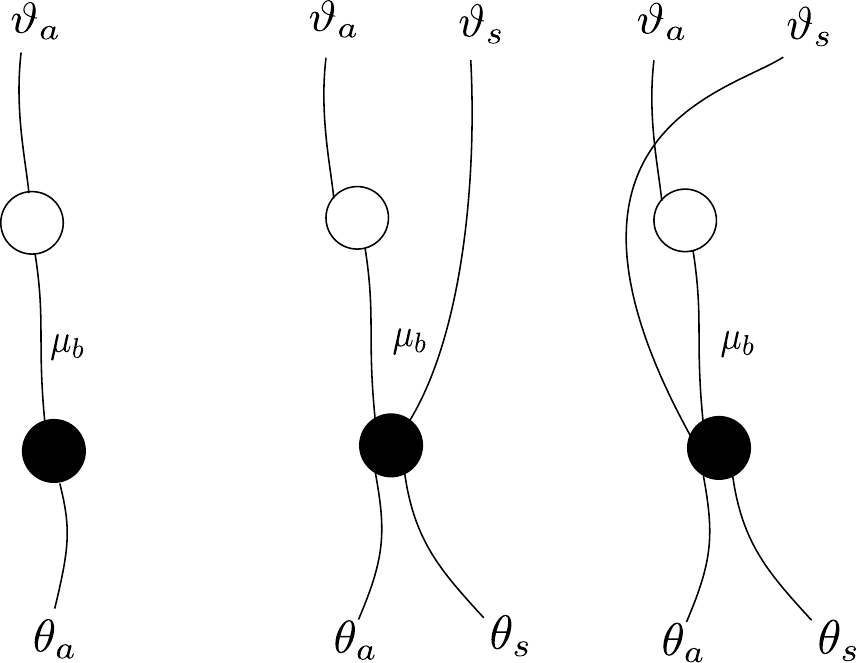}
\centering
\caption{Form-factor diagrams contributing at leading order to 1-particle (first two) and 2-particle $a\to b$ transition (last three). Secular terms arise from the second, fourth (for $\theta_s<0$) and fifth (for $\theta_s>0$) diagram.
}
\label{fig:atob}
\end{figure}

We can test the proposal through a low-temperature expansion of the finite-volume spectral sum \eqref{eq:defGR}.  Using the kinematic-pole equations and Gaudin normalization~\cite{PozsgayTakacs2008a} the complete leading contribution from one thermal spectator (second diagram in Fig.~\ref{fig:atob}) is
\begin{equation}
m
\int \frac{dy}{2\pi}\,
|\sinh(y)| e^{-\frac{m}{T}\cosh y}(c(y)-1).
\label{LO}
\end{equation}
Both its real and imaginary parts follow from the regulated calculation. Equation 
\eqref{LO} agrees with the first thermal order of \eqref{main}.
The physical interpretation is the same as Sachdev and Young's \cite{SachdevYoung}: collisions with dilute thermal particles are approximately independent and Poisson distributed. However, each collision now contributes its rapidity-dependent scattering phase \(S\), rather than universally \(-1\). 
The full integral equation relaxes the diluteness assumptions and systematically takes into account the self scattering of thermal particles. 

This first order calculation easily extends to diagonal theories with several particle species. For a transition \(a\to b\), the leading order oscillation 
is given by the third diagram of Fig. \ref{fig:atob}. There is no singularity and the integrand after momentum conservation localizes at $\mu_b=0$ giving 
\(
Z_{ab}(t)\sim t^{-1/2}e^{it(m_b-m_a)} .
\)
The damping of this oscillation comes at leading order from the two different routings on the fourth and fifth diagrams of Fig \ref{fig:atob}. The form factor singularities provide  
the spectator source with opposite signs leading to
\eq
\label{e.twopfirstorder}
\sum_s\int_{-\infty}^\infty\frac{dy}{2\pi}\,
m_s|\sinh y|\,e^{-\frac{m_s}{T}\cosh y}
\bigl[c_s^{ab}(y)-1\bigr]
\eqx
where 
\(
c_s^{ab}(y)
=
\frac{S_{sb}(-|y|)}{S_{sa}(-|y|)}.
\)
For the vacuum-to-particle transition the denominator is omitted. This determines the leading damping and frequency shifts of both the individual particle oscillations and the mixed-mass oscillations.


\begin{figure}[t]
    \centering
    \includegraphics[width=0.9\columnwidth]{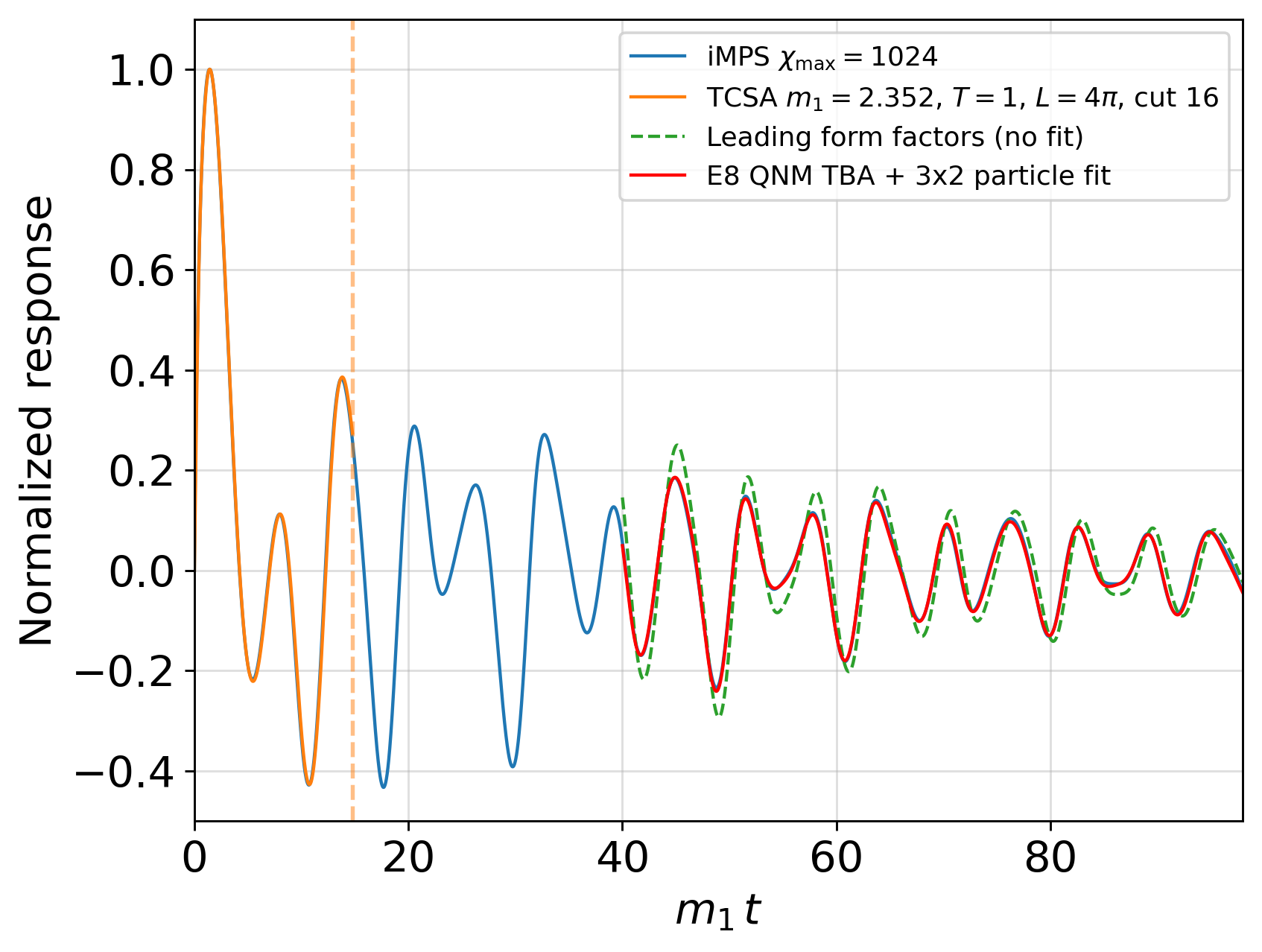}
    \caption{Tensor network simulation of the $\cor{\sg\sg}_T$ thermal response in the E8 theory, normalized to the first peak height. Comparison with a TCSA~\cite{YurovZamolodchikov1990} computation valid for small~$t$, a leading form factor prediction and a fit of the E8 QNM TBA frequencies and three 2-particle contributions (see text).}
    \label{fig.E8}
\end{figure}

In order to numerically cross-check the computations, we performed a tensor network (iMPS) simulation of the infinite volume integrated $\cor{\sg\sg}_T$ thermal retarded correlation function in the magnetic deformation of the Ising CFT (the E8 theory~\cite{Zamolodchikov1989E8}). 
In Fig.~\ref{fig.E8}, we compare the tail with a direct leading form-factor computation (exponentiating the secular term $\propto t$) using~\cite{Delfino:1995zk,Delfino:1996jr,Szecsenyi:2012jq}, as well as with a fit of the amplitudes of the damped oscillations predicted by the E8 QNM TBA supplemented by three leading 2-particle contributions resummed by a similar E8 QNM TBA with driving terms given by $c_s^{ab}$ (see \textit{Supplemental Materials} for details). The fit agrees very well with the iMPS result.


\medskip

\noindent {\bf Some additional conjectures.}  
As the conjectured proposal~\eqref{e.tba1gen}-\eqref{e.tba2gen} with sources given by~\eqref{e.sources}, provides for us a single thermal damping coefficient at $k=0$ for each particle species, it is clear that it does not yet predict the full QNM spectrum. Indeed, on the form-factor expansion side, the sketched arguments involved only a subset of secular terms, missing e.g. 2-particle contributions entering the comparison in Fig.~\ref{fig.E8}. 
Moreover, in the UV limit, we know that we should get separate sets of QNM frequencies~\eqref{e.cftqnm} associated with each primary field of the CFT. 

It is very natural to conjecture, however, that the remaining QNMs may be obtained by analytical continuation~\cite{Dorey:1996re} as in the standard excited state TBA, providing additional driving terms (see also~\cite{Bazhanov:1996aq}). Performing this procedure for the ``QNM state'' equation~\eqref{e.tba2gen}, we automatically obtain generically the correct answers in the UV limit, as the system would then directly reduce to the setup yielding~\eqref{e.Gmcft} discussed earlier.

Another very interesting direction is ``generalized thermalization'', namely the approach to a generic GHD background, in contrast to the thermal ensemble considered up till now. 
It is a reasonable expectation that the appropriate modification would just involve changing the conventional driving term
\eq
\f{m}{T} \cosh y \longrightarrow m \sum_{k \in {\mathcal I}} \beta_k \cosh ky
\eqx
However, as we do not have at our disposal the analogous conceptual structure to the thermal case which motivated our construction, this definitely requires further investigation and verification.


\medskip

\noindent {\bf Discussion.} 
In this paper, we propose that thermal damping and oscillation frequencies, combined together into complex Quasi-Normal-Mode (QNM) frequencies can be obtained through a~TBA formulation with a nonstandard analytical structure, which we call QNM TBA. This conjecture is motivated, on the one hand, by the observation that QNM frequencies appear on a complex spectral sheet extending from conventional excited state gaps on a Lorentzian cylinder of size $1/T$ with positive momentum, which clearly have a TBA description for integrable QFTs. On the other hand, direct form-factor considerations lead to the emergence of the same structure, providing, in addition, concrete driving terms for the QNM TBA.

Our proposal entails an immense simplification in comparison to conventional approaches to tackling thermal damping, which require a resummation of an infinite number of secular terms together with overcoming problems with non-convergence of form-factor expansions for timelike separations. 

The QNM TBA reproduces the exactly known damping~\eqref{e.essler1} of $\sg$ in the thermally deformed Ising CFT~\cite{Granet_2020}, has the correct low-$T$ behaviour ala Sachdev and Young~\cite{SachdevYoung}, and is consistent with iMPS tensor network simulations in the E8 theory. It is important, however, to perform further extensive numerical checks to probe and verify the detailed structure of the QNM TBA and its conjectured ``excited state'' variants in the spirit of~\cite{Dorey:1996re}.

The proposal opens up numerous further directions of research, ranging from more rigorous derivations through resummations of secular terms in form-factor expansions to investigating how the QNM TBA works in more complicated integrable QFTs and/or integrable lattice models. 
On a more general level, this unlocks an entirely new arena of thermal physics amenable to integrable TBA/Y-system methods.

On the holographic side, given the identification of CFT thermal damping with QNMs of a BTZ black hole, it is an extremely intriguing question to understand the gravitational holographic interpretation of the QNMs in the integrable massive deformations of CFTs. Holographic considerations~\cite{Janik:2025zji} also motivate future studies of similar phenomena but for these theories in finite volume and the corresponding recurrence effects. 

Last but not least, it would be very interesting to explore ``generalized thermalization'', i.e. relaxation to a Generalized Gibbs Ensemble of GHD~\cite{Castro_Alvaredo_2016,Doyon_2020}, and whether it could be described by the conjectured modification of the driving terms.

\medskip
\noindent{\bf Acknowledgements.} 
We would like to thank Benjamin Doyon, Frank Göhmann, Márton Kormos, Miłosz Panfil, Gábor Takács and ChatGPT and Claude for illuminating discussions.
We acknowledge crucial computations performed by the AI models, 
and Claude pointing out the connection of our results to ballistic fluctuation theory. 
The TCSA computation used Ising CFT data from~\cite{Horv_th_2022}.
The research was supported by Priority Research Area DigiWorld grant under the Strategic Programme Excellence Initiative at the Jagiellonian University (Kraków, Poland). RJ was supported by the National Science Centre (NCN), Poland,
Grant No.~2025/57/B/ST2/02102, while ZB by the NKKP-Advanced Grant 152467. 


\bibliography{references}

\clearpage

\appendix*
\setcounter{equation}{0}
\setcounter{figure}{0}
\renewcommand{\thefigure}{A\arabic{figure}}

\section*{Supplemental material}

\section*{Resumming the thermal two-point function}

\subsection*{The correlator and its two asymptotic regimes}

We consider a massive integrable QFT with a single particle species and a local scalar operator $O$ with a non-vanishing one-particle form factor $F_1$. The thermal two-point function is defined through the finite-volume regularisation
\begin{equation}
G(x,t)=\lim_{R\to\infty}\frac{\Tr_R\!\big[e^{-H_R/T}\,O(0,0)\,O(x,t)\big]}{\Tr_R [e^{-H_R/T}]} ,
\label{eq:Gdef}
\end{equation}
and we are interested in two complementary asymptotic regimes. At equal time and large separation
\begin{equation}
G(x,0)=\langle O\rangle_T^2+F\,e^{-x\,\Delta E}+\dots ,
\end{equation}
with $F$ sub-exponential and $\Delta E$ the thermal screening gap; at $T=0$ the one-particle contribution localises at zero momentum, $F\propto|F_1|^2/\sqrt{x}$ and $\Delta E=m$. At fixed position and large time
\begin{equation}
G(0,t)=\langle O\rangle_c^2+A\,e^{t(-\Gamma+i\Omega)}+\dots ,
\end{equation}
a damped oscillation around the asymptotic value; at $T=0$ the same one-particle integral gives, by stationary phase, $A\propto|F_1|^2/\sqrt{t}$, $\Gamma=0$ and $\Omega=m$ (plus the negative-frequency partner). The integrated correlator $R(t)=\int dx\,G(x,t)$ projects on zero momentum instead of localising on the saddle; it has the same exponent without the $t^{-1/2}$ suppression, $R(t)-R_c=\bar A\,e^{t(-\Gamma+i\Omega)}+\dots$. The two exponents, $\Delta E$ and $-\Gamma+i\Omega$, are singularities of one analytically continued propagator. Here we connect them through the form factor expansion; this is the origin of \eqref{KtoJ}--\eqref{Gammam} of the main text. Note that we consider here $-\Gamma+i\Omega$ and not the analogous $-i\,\omega_{\QNM}=-\Gamma -i\Omega$ quantity, which we introduced before, but they are simply related and the real response contains both.

\subsection*{Equal time: the mirror gap and its direct-channel origin}

\paragraph{Mirror channel.}
By the double Wick rotation the equal-time correlator is a zero-temperature, finite-volume correlator of the mirror theory on a cylinder of circumference $1/T$, with $x$ as Euclidean time. Inserting mirror eigenstates, 
\[ G(x,0)=\sum_n{}_{r}\langle0|O|n\rangle_r\,{}_r\langle n|O|0\rangle_r\,e^{-x(E_n-E_0)} \quad ,\] the large-$x$ behaviour is governed by the standing one-particle state, whose TBA is \eqref{eps1} of the main text. Taking the difference of the two TBA equations, $e^{-\varepsilon_1}=e^{-\varepsilon_0}\,s\,e^{K\star\chi}$ for $\chi=L_1-L_0$, one obtains the gap equation~\eqref{gap},
\begin{equation}
\begin{aligned}
\chi=& \log\!\big(1-f+f\,s\,e^{K\star\chi}\big) \\ 
\Delta E_1-m=&-m\!\int\!\frac{d\theta}{2\pi}\cosh\theta\,\chi(\theta),
\label{eq:gapS}
\end{aligned}
\end{equation}
with $s(\theta)=S(\theta+i\pi/2)$ in the sinh-Gordon model. 

\begin{figure}[t]
\includegraphics[width=4cm]{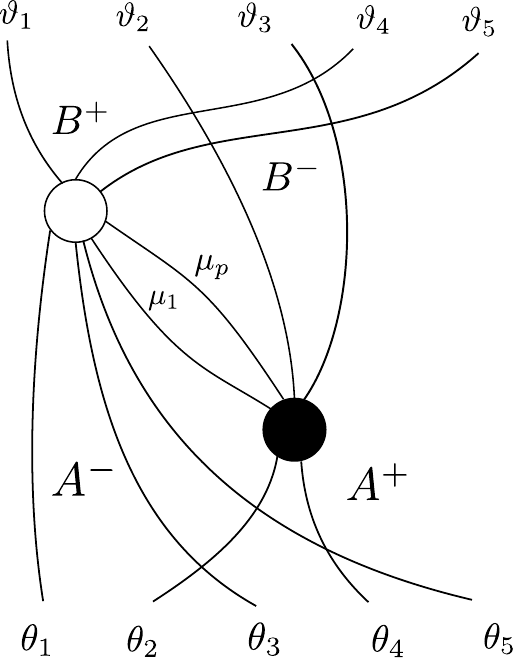}
\centering
\caption{Graphical representation of a term of the bilocal form factor \eqref{eq:PSzrepm} for five thermal particles. Crossings denote S-matrix factors; upper/lower lines attached to the circles denote outgoing/incoming particles of the respective form factor.}
\label{fig:lower}
\end{figure}

\begin{figure}[t]
\includegraphics[width=4cm]{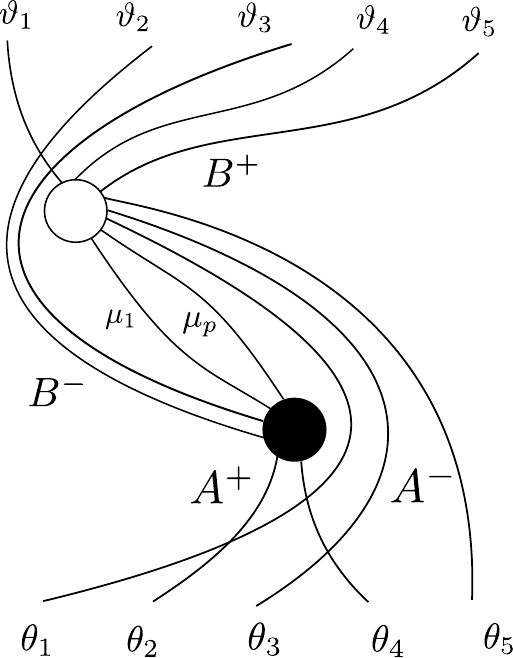}
\centering
\caption{The analogous term in the representation with the upper $\mu$-contour. What changes is the side of the $\mu$-bundle from which the particles of $B^-$ and $A^-$ approach the operators.}
\label{fig:upper}
\end{figure}

\paragraph{Direct channel.}
Inserting two resolutions of the identity in \eqref{eq:Gdef},
\begin{equation}
\begin{aligned}
G(x,t)=&\lim_{R\to\infty}\frac1{Z_R}\sum_{a,b}\langle a|O|b\rangle\langle b|O|a\rangle \times \\ 
& \qquad \qquad 
e^{-E_a/T}\,e^{it(E_b-E_a)-ix(P_b-P_a)}\, ,
\end{aligned}
\end{equation}
which holds for any $x,t$, evaluating the normalised sum before $R\to\infty$ \cite{Szecsenyi:2012jq} fixes the disconnected subtractions and the kinematical-pole prescriptions, and integrating over $x\in[0,R]$ produces $R\,\delta_{P_a,P_b}$, so that $R(t)$ keeps only the intermediate states carrying the momentum of the thermal state. For space-like separations PSz \cite{Pozsgay:2018tns} reorganised the double sum into the LeClair--Mussardo~\cite{LeClairMussardo1999} type series \eqref{PSz} of the main text, where ${\cal F}_{n,c}$ is the connected diagonal matrix element of the bilocal operator, defined as the finite part in independent regulators $\vartheta_i=\theta_i+i\epsilon_i$ of $F_n$ below. Inserting intermediate states expresses it through vacuum form factors. For $x>0$ the intermediate contours can be placed below the real axis,
\begin{equation}
\begin{aligned}
& F_n(\{\vartheta\}|\{\theta\})=\sum_{p=0}^{\infty}\frac1{p!}\!\int_{\mathbb R-i\alpha}\prod_{i=1}^{p}\frac{d\mu_i}{2\pi}
\!\!\sum_{\substack{A^+\cup A^-=\{1\dots n\}\\ B^+\cup B^-=\{1\dots n\}}}\!\!\\ 
&K_{x,t}\big(\{\mu\},\{\vartheta\}_{B^-}\big|\{\theta\}_{A^+}\big)
F\big(\{\vartheta\}_{B^+}+i\pi,\{\theta\}_{A^-},\{\mu\}_<\big)\\
&\qquad \times F\big(\{\vartheta\}_{B^-}+i\pi,\{\mu\}_>+i\pi,\{\theta\}_{A^+}\big)\\
&\qquad \times S\big(\{\theta\}_{A^-},\{\theta\}_{A^+}\big)\,S\big(\{\vartheta\}_{B^-},\{\vartheta\}_{B^+}\big),
\end{aligned}
\label{eq:PSzrepm}
\end{equation}
with $A^\pm$ ordered increasingly, $B^\pm$ decreasingly, the ordering of the $\mu$'s indicated by the subscript, the external S-matrix phases collected in the last two factors, and
\begin{equation}
K_{x,t}\big(\{\alpha\}\big|\{\beta\}\big)=e^{it\left(E(\{\alpha_j\})-E(\{\beta_k\})\right)}e^{-ix\left(p(\{\alpha_j\})-p(\{\beta_k\})\right)}
\end{equation}
and $p=m\sinh,\ E=m\cosh$.  A term of \eqref{eq:PSzrepm} is represented graphically in Fig.~\ref{fig:lower}. 
For $x<0$ the contours have to be placed above the real axis. Shifting them across the kinematical poles at $\mu_j=\theta_i$ and $\mu_j=\vartheta_i$, collecting the residues and regrouping them with the permutation and kinematical axioms, PSz obtained the equivalent representation of Fig.~\ref{fig:upper}: the particles of $B^-$ and $A^-$ approach the operators from the other side of the $\mu$-bundle, with correspondingly reordered form factor arguments and S-matrix phases.

\paragraph{Secular terms and the connected logarithm.}
To keep track of the thermal order we introduce a counting parameter $z$ through $f(\theta)\mapsto z\,f(\theta)$ 
and expand the gap and the amplitude,
$\Delta E_1=m+z\,\delta M_1+z^2\delta M_2+\dots$, $A_T/A_0=1+z\,a_1+z^2a_2+\dots$. The normalised connected correlator then reads
\begin{equation}
\begin{aligned}
&\frac{G(x,0)_c}{A_0e^{-mx}}=1+z\,(a_1-x\,\delta M_1)\\
&\hspace{0.2cm}+z^2\Big[a_2-x\,(\delta M_2+a_1\delta M_1)+\tfrac{x^2}{2}(\delta M_1)^2\Big]+\dots
\label{eq:zexp}
\end{aligned}
\end{equation}
The coefficient of $x$ in the connected logarithm therefore determines the gap and the $x$-independent terms the amplitude, and the second correction $\delta M_2$ can only be isolated once the first amplitude correction $a_1$ is known. In the series \eqref{PSz} the secular ($x$-polynomial) terms originate from the kinematical poles, both through the regulators (the $\epsilon$-dependence of $K_{x,t}$) and through the integration domains. 
In terms of the thermal filling fraction  one finds to second order
\begin{equation}
\chi_1=f\,(s-1)\ ,\ \ 
\chi_2=f\,s\,K\star\!\big[f\,(s-1)\big]-\tfrac12\,f^2(s-1)^2 ,
\label{eq:chi12}
\end{equation}
in agreement with the expansion of \eqref{eq:gapS}: the first term in $\chi_2$ is the
self-scattering of the thermal particles with the propagating one through
the kernel, the second the exclusion of coinciding rapidities. 
Although not established to all orders, the mirror representation guarantees that the $x$-linear part of the connected logarithm is resummed by \eqref{eq:gapS}. We use this fact to obtain the time-like analogue. The explicit terms of \eqref{eq:zexp} up to third order can be found in \cite{Bajnok:2018fzx,Bajnok:2023bpj,Bajnok:2019cdf}.

\paragraph{Which configurations carry the exponent.}
For $x>0$ the lower contours can be pushed down to $\Imo(\mu)=-\pi/2$ without crossing singularities (no bound states). Each intermediate particle then carries $e^{-xm\cosh\Reo(\mu)}$, so terms with $p$ intermediate particles belong to the $p$-particle line $e^{-pmx}$ and the leading exponent comes from $p=1$, with the $\mu$-integral localised at $\Reo(\mu)=0$, and from thermal particles whose two legs are attached to the same operator (diagonal configurations). Thermal particles attached to different operators carry phases $e^{\pm ixp(\theta_i)}$; after the $\theta_i$-integration and localization they contribute to the finite-volume matrix element, i.e.\ the prefactor. Configurations in which a thermal particle continues as an intermediate one, $\theta_i\to\mu_j\to\vartheta_i$, have a double pole on the convergent side of the contour,  the near-pole region of the $\mu_j$-integral vanishes identically and the remainder is exponentially suppressed. In this bookkeeping the source arises from the orientation of the S-matrix between a thermal particle and the propagating one: the lower routing produces $S(\theta-u)$, which after $u\to u-i\pi/2$ and localisation at $u=0$ becomes $S(\theta+i\pi/2)=s(\theta)$, while the diagonal thermal particles generate the finite-volume form factor through their regular parts and the exponential through the $\epsilon$-derivative of $K_{x,t}$ at their poles.

\subsection*{Time-like separations: light cone, crossed poles and rerouting}

\begin{figure}[t]
\centering
\includegraphics[width=8cm]{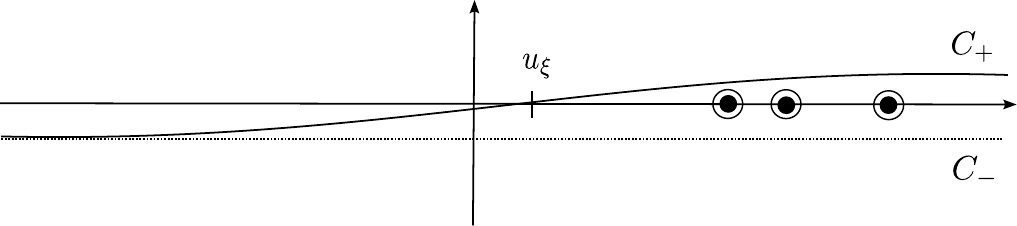}
\caption{Contour deformation for time-like separations. The lower contour is kept for $u<u_\xi$ and moved above the real axis for $u>u_\xi$; the kinematical residues of the crossed poles have to be picked up.}
\label{fig:contour}
\end{figure}

\begin{figure}[t]
\centering
\includegraphics[width=4cm]{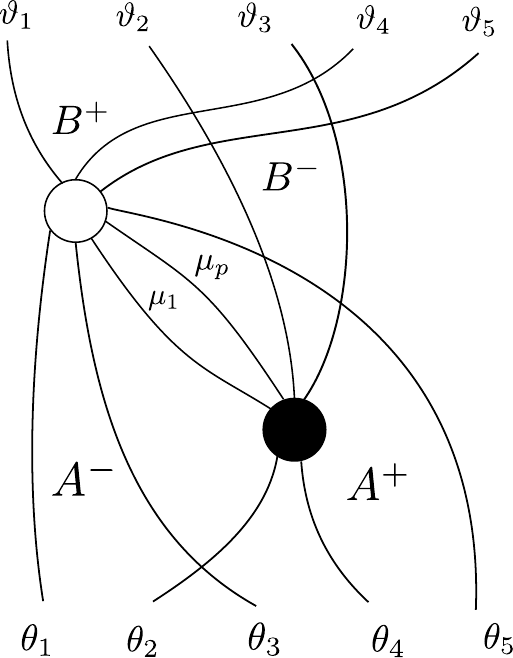}
\caption{Time-like prescription in the integration domain in which only $\theta_5$ is larger than $u_\xi$: the particle $5$ uses the upper routing, all others the lower one. This leads to additional S-matrix elements with $\vartheta_2$ and $\vartheta_3$.}
\label{fig:hybrid}
\end{figure}

Let us now switch on $t>0$ and introduce $\xi=x/t$. A propagating rapidity $u$ carries the phase
\begin{equation}
e^{it\left[E(u)-\xi p(u)\right]},\qquad \kappa_\xi(u)=p(u)-\xi E(u),
\end{equation}
and under a local shift $u\mapsto u+i\eta$ the real part of the exponent is $-t\sin\eta\,\kappa_\xi(u)$. For space-like separations, $|x|>t$, $\kappa_\xi$ has a definite sign on the real axis and a uniform deformation is allowed. For $t>|x|$ it changes sign at
\begin{equation}
u_\xi=\operatorname{artanh}\xi\ ,\ \  \sigma_\xi(u)=\sgn\kappa_\xi(u)=\sgn(u-u_\xi),
\end{equation}
so the convergent contour passes below the real axis for $u<u_\xi$ and above it for $u>u_\xi$, joined at $u_\xi$ (Fig.~\ref{fig:contour}). Moving the part $u>u_\xi$ of the lower contour of \eqref{eq:PSzrepm} upwards is the PSz contour shift restricted to the poles above $u_\xi$: one picks up the residues of the integrand on the real line,
\begin{equation}
\int_{C_-}\frac{du}{2\pi}F(u)=\int_{C_+}\frac{du}{2\pi}F(u)+i\!\!\sum_{\text{crossed poles}}\!\!\operatorname{Res}F ,
\end{equation}
evaluated with the annihilation residue
\begin{equation}
-i\operatorname{Res}_{v=u}F_{n+2}(v+i\pi,u,\boldsymbol\alpha)=\Big[1-\prod_{j=1}^nS(u-\alpha_j)\Big]F_n(\boldsymbol\alpha).
\label{eq:resid}
\end{equation}
Iterating \eqref{eq:resid} generates multiple residues and changes both the contours and the integrands, but the PSz regrouping is iterative in the particle number and applies directly to a subset of poles. The outcome is a representation with a rapidity-dependent routing: thermal particles with $\theta_i<u_\xi$ keep the lower routing of Fig.~\ref{fig:lower}, those with $\theta_i>u_\xi$ acquire the upper routing of Fig.~\ref{fig:upper}, and the only new elements are S-matrix factors between particles on opposite sides of $u_\xi$ (see Fig.~\ref{fig:hybrid}), which 
are absent
in the two uniform cases. The reduction of the previous section carries over: the propagating $\mu$ localises at the real saddle $u_\xi$ (for the integrated correlator $I(t)$, it is pinned there by momentum conservation), terms with more intermediate particles oscillate at other frequencies, continuation configurations again contribute nothing to the exponent on the convergent side, and the exponent of the one-particle line is determined by diagonal thermal particles and a single $\mu$, which we consider from now on.

\subsection*{Local transformation of the scattering kernel and the source}

We now compare the regulated residues before and after the partial deformation, focusing on $\xi=0$, $u_\xi=0$, which is the case relevant for the integrated correlator. For an integrand with $n$ thermal particles the integration domain splits into $2^n$ sectors according to the signs of the $\theta_i$, each with its own routing. In the connected part the product of form factors has the generic singularity $\prod_i\epsilon_i^{-1}$ and one has to pick, in the numerator, the term proportional to $\prod_i\epsilon_i$. We separate this numerator into a universal part, built from $K_{x,t}$ and S-matrices only, and the regular parts of the form factors. The latter are operator dependent and contribute to the prefactor and to the lower-order terms in \eqref{eq:zexp}, the former determine the exponent. Introduce
\begin{equation}
\sigma_i=\sgn\theta_i\ ,\ \   c_i=S(-|\theta_i|)\ ,\ a_i=-m\,t\,\sinh\theta_i=-\sigma_i|a_i| ,
\end{equation}
and $K_{jk}=K(\theta_j-\theta_k)$. Then the universal term is the part of
\begin{equation}
\;\prod_{j=1}^{n}e^{\epsilon_ja_j}\,\rho_j(\{\epsilon\})\ ,\ \ 
\rho_j(\{\epsilon\})=-\sigma_j\Big(c_j\,e^{-\sigma_j\sum_{k\neq j}\epsilon_kK_{jk}}-1\Big)\;
\label{eq:effnum}
\end{equation}
proportional to $\prod_i\epsilon_i$.
This is an all-sector statement. The ingredients are the following. (i) For a thermal particle whose two legs are attached to the same operator the kinematical pole has the residue \eqref{eq:resid}, with the product running over the other legs of that operator: it contains the S-matrix with the propagating particle, $S(\theta_j-u)$ in the lower and $S(u-\theta_j)$ in the upper routing, which at $u=0$ is $c_j$ in both cases, and for every other thermal particle $k$ attached to the same operator the pair $S(\theta_j-\vartheta_k-i\pi)S(\theta_j-\theta_k)=e^{-\epsilon_kK_{jk}}$. (ii) The kinematical factor contributes $e^{\epsilon_ja_j}$ when both legs of~$j$ sit on $O(x,t)$ and nothing otherwise; attaching both legs to $O(0,0)$ reverses the sign of the residue, which is what cancels the poles. (iii) The regulator dependence of the rerouting phases supplements (i) in such a way that, after summing over the two attachments of every particle, the coefficient of $\epsilon_k$ carried by particle $j$ is $K_{jk}c_j$ in every sector; the distribution of this coefficient between residues and phases differs from sector to sector, but the sum does not. The orientation thus enters only through $\sigma_j$, in $a_j$, in the overall sign of $\rho_j$ and in its exponent. For a uniform sign, $\sigma_j\equiv\sigma$, \eqref{eq:effnum} reduces to the space-like expression (the two values of $\sigma$ corresponding to the two PSz representations, with $c\to s$ after the rotation of $u$). For $\sigma\equiv+1$ the phases carry no regulators and \eqref{eq:effnum} follows to all orders from the nested structure of the residues, each residue being exact in the regulators not yet removed; for $\sigma\equiv-1$ the same sum is redistributed between residues and phases, which is the prototype of the mixed-sector mechanism (iii).

At first thermal order \eqref{eq:effnum} gives $a_j\rho_j=|a_j|(c_j-1)$, i.e.\ the QNM frequency \eqref{LO} of the main text with its real and imaginary parts. The same first-order mechanism in a diagonal theory with several species gives \eqref{e.twopfirstorder} with the spectator source $c^{ab}_s$.
This also leads to the time-like counterpart of $\chi_1$ in~\eqref{eq:chi12}, 
\eq
\chi_1=f(c-1)
\eqx
Going on to second thermal order, \eqref{eq:effnum} directly generates just the first term of $\chi_2$ of~\eqref{eq:chi12} with the replaced symbols $s\to c$ and $K\to J=\sg K \sg$:
\eq
f\,c\,J\star\!\big[f\,(c-1)\big]
\eqx
The second term of $\chi_2$ comes from a ``contact term'' when thermal rapidities coincide (indeed it is proportional to $\chi_1^2$). In the uniform sectors, corresponding to space-like separation, this process of combining direct contributions and contact terms should continue (with $s$ and $K$) and effectively transform
\eq
f\left(s e^{K\star \chi} - 1\right)  \longrightarrow \log\left(1 + f\left(s e^{K\star \chi} - 1\right) \right)
\eqx
in order to be equivalent to a resummation by the gap equation~\eqref{eq:gapS}. 
This is somewhat reminiscent of the effects of ``multiple wrapping'' in~\cite{Bajnok:2017mdf}.

We expect that within the mixed sectors, i.e. for time-like separation, the direct contribution of~\eqref{eq:effnum} will amount to the substitution $s\to c$ and $K\to J=\sg K \sg$, as we demonstrated in the first two orders.  Since the combinatorics of the coinciding thermal rapidities will act in an identical manner, 
 we can inherit the resummation from the space-like case and in this way arrive 
at the gap equation~\eqref{main} conjectured in the main text:
\eq
\chi=\log \left(1-f+f\,c\,e^{J\star\chi}\right)
\eqx
The resummed direct secular terms proportional to $t$ are coming from one single $a_j$ term and all possible contributions from its $\rho_j$ (all together proportional to $\prod_i \epsilon_i$).  They can be represented as trees with a root at $a_j$, where  oriented links are given by $\sigma_j K_{jk} $, vertices carry factors $-\sigma_j c_j$ and leaves $-\sigma_k(c_k-1)$. This structure is resummed by $f(se^{K\star \chi}-1)$ which is then supplemented by the contribution of coinciding rapidities to give 
\eq
t\,m\!\int\!\frac{d\theta}{2\pi}\,|\sinh\theta|\,\chi(\theta)
\eqx
which leads to the damping formula~\eqref{Gammam} in the main text.
To summarize, 
the only
effect of the time-like kinematics is $K\to J=\Sigma K\Sigma$, $s\to c$ and
$m\cosh\theta\to m|\sinh\theta|$, i.e.~\eqref{KtoJ}.

The rerouting is the direct-channel counterpart of the $i\pi$ shift of the right half-line in the QNM TBA, Fig.~\ref{fig.contours}(left) of the main text: $K(y_1-y_2+i\pi)=-K(y_1-y_2)$ turns $K$ into $J$ and $\cosh(y+i\pi)=-\cosh y$ produces the sign in the driving term \eqref{e.drivingtheta}. 

Since the resummation itself is inherited from the space-like case, where the mirror channel makes it exact, the only ingredient specific to time-like separations is the local form \eqref{eq:effnum} of the universal numerator in the mixed sectors, established here to second thermal order.

\subsection*{Connection to ballistic fluctuation theory }

The excited state TBA equation in the thermal channel has a transparent meaning. The standing particle of the finite-volume channel acts as a defect in the thermal gas \cite{Bajnok:2007jg}: inserted into the thermal trace it is the defect operator of \cite{Bajnok:2004jd}, which acts on multiparticle states by multiplying each particle with its transmission factor, i.e. it acts as an additional conserved charge with one-particle eigenvalue $h(\theta)=\log S(\theta+\tfrac{i\pi}{2})=\log
s(\theta)$, and the gap is the generating function of this charge in the interval between the two equal-time operators. For operators at equal position the Lorentz-rotated quantity measures a time-integrated current instead: the net flux of the same charge through the worldline connecting the two operators, with $h(\theta)=\log S(-\theta)$ and the
direction of crossing accounted for by the sign $\Sigma$. Ballistic fluctuation theory \cite{Doyon:2019,Myers:2020} determines how the insertion of $e^{\lambda\int_0^t j\,dt'}$ modifies the generalized Gibbs ensemble through the flow equation
$\partial_\lambda\varepsilon=-\mathrm{sign}(v^{\rm eff})\,h^{\rm dr}$, where $v^{\rm eff}$ is the effective velocity of the quasiparticles and $h^{\rm dr}=h+K\star(n\,h^{\rm dr})$ the thermally dressed charge; the exponent is then the integral over $\lambda$ of the current expectation value. For the integrated correlator the sign is fixed by parity to $\Sigma$, and since $n\,h^{\rm dr}=\Sigma\,\partial_\lambda L_\lambda$, pulling the sign through the dressing turns $K$ into $J=\Sigma K\Sigma$,
$\partial_\lambda\varepsilon=-\Sigma h-J\star\partial_\lambda L_\lambda$. Integrating from $\lambda=0$ to $1$ with $\Sigma h=\log c$ gives \eqref{e.sources} and hence \eqref{main}. 
In the same language the gap \eqref{E1} is the change of the free energy, $-\int\frac{dp}{2\pi}(L_1-L_0)$, and the damping \eqref{Gammam} the change of the free energy flux, $\int\frac{|dE|}{2\pi}(L_1-L_0)$, produced by the same twist. These are the two potentials of   generating charges and currents respectively \cite{Bajnok:2019}, with $|dE|$ because the line is crossed in both directions. For a constant twist this structure, including the sign of the effective velocity, is the one found for the topological charge of the sine-Gordon model in~\cite{DelVecchioDelVecchio:2023}.

\section*{Tensor network (iMPS) simulations of the E8 theory}

We realize the E8 theory, i.e. the magnetic deformation of the Ising CFT through its spin chain construction~\cite{KjallPollmannMoore2011}
\eq
\label{e.isingspinchain}
H=-J\sum_i \sigma_i^x\sigma_{i+1}^x-J\sum_i\sigma_i^z + h_\sg \sum_i \sigma_i^x
\eqx
We take $J=1$ and the lattice spacing $a=1$. For $h_\sg \to 0$, as is necessary for the continuum limit of the theory, the lattice velocity $v_{lat}=2$. Then the continuum time, temperature and mass are related to the lattice ones through
\eq
t_{cont}=\frac{v_{\rm lat}}{a}t_{lat},
\quad
T_{cont}=\frac{a}{v_{\rm lat}}T_{lat},
\quad
m_{cont}=\frac{a}{v_{\rm lat}}\Delta_{lat}
\eqx
where $\Delta_{lat}$ is the energy gap in lattice units.

We use the \texttt{tenpy} library~\cite{HauschildPollmann2018} for the tensor network simulations. All the computations are done directly in infinite volume. First, a VUMPS~\cite{ZaunerStauber2018} is constructed for the ground state and the dispersion of plane wave excitations is computed~\cite{Haegeman2012}. From the value at $k=0$ we obtain $\Delta_{lat}$ and from the coefficient of $k^2$ we confirm that $v_{lat} \simeq 2$. For $h_\sg = 0.002$, we get $m_{cont}=0.098$ and $v_{lat}=2.009$. $m_{cont}$ is identified with the lowest particle mass $m_1$ in the E8 theory, which we denote below by $m$.

The thermal density matrix is represented by an infinite \texttt{PurificationMPS} (iMPS)~\cite{VerstraeteGarciaRipollCirac2004,FeiguinWhite2005} with cell width~2, obtained from the infinite temperature one by MPO evolution in imaginary time to $\beta_{lat}=12$. We use bond dimension $\chi=64$ for this stage of the computation. We construct the thermal state with $m/T=2.352$.

The integrated retarded correlator is computed using its interpretation through a pertubation of the thermal density matrix as in~\cite{Janik:2025zji}.
Introducing the spatially integrated operator
\eq
O(t) =  \int_{-\infty}^{\infty} \f{dx}{2\pi} \OO{t,x}
\eqx
we perturb the thermal density matrix at $t=0$ by acting with $\exp(i \epsilon\, O(0))$ and focus on the term at linear order in $\epsilon$:
\eq
\rho_T \to e^{-i \epsilon\, O(0)} \rho_T e^{i \epsilon\, O(0)} \sim 
\rho_T + \epsilon \,\underbrace{i  [\rho_T,\, O(0)]}_{\dl\rho} + \ldots 
\eqx
Evolution of the perturbation $\dl\rho$ in time and measurement of the expectation value of the $\OO{t,0}$ operator yields
\eq
\label{e.tcsaobs}
\cor{O(t)}_{\dl\rho(t)}\!\! = \tr O(t)\dl\rho(t) = 
\tr O(t) e^{-i H t} \dl\rho\, e^{i H t}
\eqx
which is equivalent to~\eqref{e.retcorr}.

As the linear perturbation $i [\rho_T,\, O(0)]$ cannot be simply represented by a \texttt{PurificationMPS} which then would be evolved in a standard way, we perform two finite but small perturbations
\eq
\rho_\pm(0)
=
e^{\mp i\epsilon O}\rho_T e^{\pm i\epsilon O}
=
\rho_T \pm \epsilon\, i[\rho_T,O] + O(\epsilon^2)
\eqx
then evolve the two infinite \texttt{PurificationMPS} states and compute
\eq
\frac{\tr[\rho_+(t)O]-\tr[\rho_-(t)O]}{2\epsilon}
=
\tr[\delta\rho(t)O] + O(\epsilon^2)
\eqx
In our case, $O$ is the integrated $\sg$ operator, which is represented on the spin chain~\eqref{e.isingspinchain} by $\sum_i \sg_i^x$.

For this part of the computation, the code from the \texttt{tenpy} library necessary for running the real-time evolution was extracted by \texttt{ChatGPT} and converted from \texttt{numpy} to \texttt{cupy} so as to run on a GPU. 
The final simulation was run up to $t=1000$ (i.e. $m\,t=98$) on an Nvidia V100 GPU, with the maximum bond dimension being $\chi=1024$ and the \texttt{backwards} disentangler~\cite{Hauschild2018purification}. 

In order to cross-check both the construction of the thermal state, as well as the tensor network evolution code, we compare the results with an independent \emph{finite volume} TCSA~\cite{YurovZamolodchikov1990} code for computing~\eqref{e.tcsaobs}. The TCSA computation was performed for the E8 theory with mass gap $m=2.352$, $T=1.0$ on a cylinder of size $L=4\pi$. We used the Ising CFT data provided in the paper~\cite{Horv_th_2022} and performed the calculation for chiral cuts $N_L, N_R\leq 16$ (comprising 59646 states) on a GPU. In Fig.~\ref{fig.E8}, we compare the TCSA evolution for $t\leq 2\pi$ (before the wrapping threshold) with the tensor network simulation and find excellent agreement.

\section*{Comparison with the E8 tensor network (iMPS) computation}

We compare the tail of the response function in the range $40 \leq m t \leq 98$ with a direct computation of the amplitudes, oscillation frequencies and thermal dampings in terms of E8 form-factors at leading order, as well as with a fit based on the proposed E8 QNM TBA.

The first order form-factor expression has the form
\eqn
R_{FF}(t)
&={}&
\sum_{a=1}^{8}
\frac{{F_a^\sg}^2}{m_a}
e^{-\Gm_a^{(1)}t}
\sin\!\left[(m_a+\delta m_a^{(1)})t\right]\nonumber\\
&+&
\sum_{(a,b)\in\mathcal P}
\operatorname{Im}\!\left[
Z_{ab}(t)
e^{(-\Gm_{ab}^{(1)}+i\delta m_{ab}^{(1)})t}
\right]
\eqnx
In Fig.~\ref{fig.E8} we included $a,b \in \mathcal{P}=\{(1,2), (1,3), (2,3)\}$, which are dominant.
The first order damping and frequency are obtained by exponentiating the secular term proportional to $t$. Concretely, we have
\eqn
&&-\Gm_a^{(1)} + i \delta m_a^{(1)} = \\
&&\ \ =\sum_{s=1}^8 \int \f{dy}{2\pi} |m_s \sinh y| e^{-\f{m_s}{T}\cosh y} \left( S_{sa}(-|y|) -1 \right)
\nonumber
\eqnx
for single particle terms and
\eqn
&&-\Gm_{ab}^{(1)} + i \delta m_{ab}^{(1)} = \\
&&\ \ =\sum_{s=1}^8 \int \f{dy}{2\pi} |m_s \sinh y| e^{-\f{m_s}{T}\cosh y} \left( \frac{S_{sb}(-|y|)}{S_{sa}(-|y|)} -1 \right)
\nonumber
\eqnx
for the 2-particle terms, where $\Om_{ab}^{(1)} = m_b - m_a + \delta m_{ab}^{(1)}$. The $F_a^\sg$ are the known 1-particle form factors of the $\sg$ operator in the E8 theory~\cite{Zamolodchikov1989E8}.

The time-dependent prefactor of the 2-particle terms is given by the integral
\eq
Z_{ab}(t)
=
\int_0^{\infty}\frac{dp}{\pi}\,
W_{ab}(p)
\left[e^{-\f{E_a(p)}{T} }-e^{-\f{E_b(p)}{T}}\right]
e^{\,i[E_b(p)-E_a(p)]t}
\label{e.Zab}
\eqx
where
\eq
W_{ab}(p) = \frac{\left[F^\sg_{ab}(i\pi+\eta_{ab}(p))\right]^2}{E_a(p)E_b(p)}
\eqx
with $F^\sg_{ab}$ being the 2-particle form factor of $\sg$, 
$E_a(p)$ is the energy of particle $a$, while $\eta_{ab}(p)=\th_a(p)-\th_b(p)$ is the difference of rapidities of the two particles.
For the QNM TBA fit described below, it is important to note that applying stationary phase approximation to~\eqref{e.Zab}, gives an overall $1/\sqrt{t}$ behaviour.
Expressions for the E8 form-factors of the $\sg$ operator are given in~\cite{Delfino:1995zk,Delfino:1996jr}.

The QNM TBA has as its source term $-\log c$ with $c(y)=S(-|y|)$. When generalized to a multiparticle diagonal scattering theory like E8, the driving term for the $s$'th component equation is $-\log c^{a}_s(y)$ with
\eq
\label{e.1pdriving}
c^{a}_s(y) = S_{sa}(-|y|)
\eqx
Here, $a=1\ldots 8$ is the fixed particle species, whose thermal shift of the oscillation frequency and damping we are computing. 
Based on the first order computation corresponding to the transition $a \to b$~\eqref{e.twopfirstorder}, it is natural to expect that the resummed QNM TBA version would have the driving term
\eq
\label{e.2pdriving}
c_s^{ab}(y)
=
\frac{S_{sb}(-|y|)}{S_{sa}(-|y|)}
\eqx
for given $a$ and $b$.

For the fit, we compute the oscillation frequencies $\Om_a$ and thermal dampings $\Gm_a$ for $a=1\ldots 8$ by numerically solving the E8 QNM TBA with $m/T=2.352$ with driving terms~\eqref{e.1pdriving}.
We supplement these by the expected three leading 2-particle contributions, whose complex frequencies $-i\Gm_{ab}+\Om_{ab}$ are computed by solving the QNM TBA with driving terms~\eqref{e.2pdriving}.
These 2-particle damped oscillations have in addition a $1/\sqrt{t}$ prefactor, which we deduce from the complete first order computation.
We thus fit the following function
\eq
Re\left(\sum_{a=1}^8 C_a\, e^{(-\Gm_a+i\Om_a)t} + t^{-\f{1}{2}} \sum_{a,b\in \mathcal{P}} C_{ab}\, e^{(-\Gm_{ab}+i\Om_{ab})t} \right)
\eqx
where $\mathcal{P}=\{(1,2), (1,3), (2,3)\}$ as above, determining 11 complex amplitudes $C_a$ and $C_{ab}$.

\end{document}